\documentclass{vgtc}                          

\let\ifpdf\relax

\usepackage{mathptmx}
\usepackage{graphicx}
\usepackage{times}
\usepackage{subcaption}
\usepackage{array}

\usepackage{listings}
\usepackage{float}
\floatstyle{ruled}
\newfloat{mylisting}{htbp}{loa}
\floatname{mylisting}{Listing}

\usepackage[none]{hyphenat}

\lstdefinelanguage{Yarp}{%
morekeywords={%
yarp, admin, rpc, prop, set, sched, policy,%
priority, qos
},%
sensitive=true,%
morecomment=[l]{--},%
morecomment=[n]{--[[}{]]},%
morestring=[b]",%
morestring=[b]',%
}[keywords,comments,strings]%

\onlineid{0}

\vgtccategory{Research}

\vgtcinsertpkg

\title{Communication channel prioritization in a publish-subscribe architecture}

\author{Ali Paikan\thanks{e-mail: ali.paikan@iit.it} %
\and Daniele Domenichelli\thanks{e-mail: daniele.domenichelli@iit.it} %
\and Lorenzo Natale\thanks{e-mail: lorenzo.natale@iit.it}}
\affiliation{\scriptsize iCub Facility\\ Istituto Italiano di Tecnologia}

\abstract{ Real-Time communication is important in distributed applications when timing constraints on task execution and data processing play a fundamental role. Software engineering does not yet specify how real-time properties should be integrated into a publish/subscribe middleware. This article describes an approach for integration of priority Quality of Service (QoS) in a publish/subscribe middleware. The basic idea is to leverage the operating system functionalities to provide a framework where specific communication channels can be prioritized at run-time. This paper presents an implementation of our approach in the YARP (Yet Another Robot Platform) open source middleware and a preliminary experimental evaluation of its performance.
} 

\CCScatlist{ 
  \CCScat{}{Real-time system}{}{};
  \CCScat{}{Quality Of Service}{}{}
  \CCScat{}{Channel prioritization}{}{};
  \CCScat{}{Publish-Subscribe Architecture}{}{};
  \CCScat{}{Communication}{}{}
}

\begin{document}

\firstsection{Introduction}

\maketitle

Concurrent execution of software components on a cluster of computer is a widely adopted solution to implement parallel computing in various fields, as for example robotics~\cite{yarp07,Quigley09} and Virtual Reality~\cite{raffin2006,rehfeld2013}. Component--based software engineering (CBSE) in distributed systems often strive to decouple communication concerns from the implementation of the components so that the latter can operate independently with respect to time, space or information flow. Several paradigms have been proposed to achieve different level of separation such as tuple spaces, message passing and Remote Procedure Calls (RPCs). In this context publish/subscribe~\cite{eugster2003many} is a paradigm that is getting increasingly popular because it allows achieving a good level of component decoupling and reusability.

In a publish/subscribe model, a sender (known as publisher) does not send events (which are either messages or remote procedure calls) directly to a specific receiver (known as subscriber). Instead, a publisher registers itself into a central Event Service as an entity that can provide specific events (characterized by a type or a simpler identifier). In an asynchronous way, subscribers express interests and receive one or more of the available events without any knowledge of the number and identity of publishers. This enables three levels of decoupling: \emph{i)} anonymity, as the publisher and subscriber are unaware about the existence of each other; \emph{ii)} time decoupling as the publisher and subscriber do not necessarily need to be active at the same time and \emph{iii)} asynchronism, as the subscriber does not need to pull events nor does it wait for delivery notification from subscribers. Overall this allows achieving a level of decoupling that has pushed adoption of the publish/subscribe paradigm in domains such as financial, automation, transportation and robotics.  

Time-critical, distributed applications require support for real-time and Quality of Service (QoS) communication. There are various architectures which aim to provide publish/subscribe in QoS--enabled component middleware: RTSE~\cite{harrison1997design}, RTNS\cite{gore2004design} and DDS~\cite{pardo2003omg}. Deng et al.~\cite{DengGanand2007} provides a comprehensive survey of these architectures and describes different design choices for implementing real--time publish/subscribe services. QoS and real--time properties can be configured at three different levels, i.e., event, port and channel level. The actual implementation, however, depends on the point in the middleware in which a service is integrated. For example, for the CORBA middleware~\cite{CORBA08}, the different choices can be the component itself, the container or the component server.%

\begin{figure}[t]
   \begin{center}
     \includegraphics[width=3.4in]{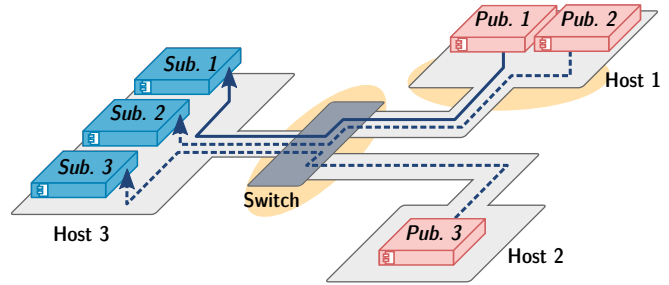}
     \caption{An example of publish/subscribe architecture in a LAN network consisting of three nodes (Host1, Host2 and Host3) connected through an Ethernet switch. The arrows represent data--flow from publishers to subscribers.}
     \label{fig:bottlenecks}
   \end{center}
\end{figure}

Different QoS can be configured into a publish/subscribe middleware to meet the requirements of real--time applications. One type of QoS is message prioritization. This allows events that require immediate reaction from a specific subscriber to be delivered in a prioritized manner. Such \emph{channel prioritization} is important in time--critical applications in fields like control systems and robotics where multiple components subscribe to a specific event (called a \emph{topic}) but only one, or some, of them need to access the messages with the lowest latency and certain guarantees. When the underlying network infrastructure is loaded by different streams of messages the priority QoS can be applied to  guarantee delivery of certain messages within the shortest time. Priorities affect messages that are in transit, that is, that have not been processed yet. This implies that two subscribers listening to the same topics may process messages in different orders even though they are properly prioritized within the communication channels. Thus, prioritization is considered as a best--effort QoS~\cite{eugster2003many}.  

This paper presents the integration of priority QoS into the YARP~\cite{yarp07} (Yet Another Robot Platform) middleware, an open source publish/subscribe middleware designed and developed for robotics. In our approach we extend the properties of individual connection channels with a priority level. This priority level affects both the priority of threads that handle the communication and the network packet's type of service. The advantage of our approach is that it relies on the functionalities offered by the operating system, it does not require specific components for message prioritization and it does not add any overhead to the communication channels. In addition and, more importantly, it allows for remote configuration of QoS at port scope and for run-time, dynamic prioritization of communication channels. This work was inspired by the requirements of robotics applications, but is indeed applicable in all time-critical, distributed applications. In the field of Virtual Reality for example the importance of reducing latencies is discussed in~\cite{rehfeld2013} to benefit synchronization and~\cite{ohl2013} for distributed acquisition and rendering with multiple cameras. 

The rest of the paper is organized as follows. Section~\ref{prioritization} describes the common problems of channel prioritization and explains how it can be integrated in a publish/subscribe middleware. The actual implementation of channel prioritization in the YARP middleware and its experimental evaluation are presented in Section~\ref{implementation}. Finally Section~\ref{conclusions} presents the conclusions and future work.

\section{Communication channel prioritization}
\label{prioritization}
A way to implement a publish/subscribe system is by using an intermediate broker to which publishers post messages. The broker then performs a store-and-forward function to deliver messages to subscribers that are registered with it. To implement message prioritization in such scenario the broker can simply route messages with a desired order. For example a subscriber to a specific topic may need to receive a copy of a message always before another subscriber. The centralized approach, however, becomes easily inefficient and does not scale well. A more efficient approach is to let components share meta--data (describing for example the type of messages) and establish peer-to-peer connections between publishers and subscribers. For example, Data Distribution Service (DDS)~\cite{pardo2003omg} uses IP multicast, YARP~\cite{yarp07} and ROS~\cite{Quigley09} use a centralized name server for storing meta--data and perform naming lookup (the approach implemented in YARP will be described in the following section). Achieving message prioritization in such distributed systems is more difficult because there is no central authority that can control message delivery. In this paper we therefore a different approach that we called ``channel prioritization''.

To better understand what we mean with ``channel prioritization'', consider the network architecture depicted in Fig.~\ref{fig:bottlenecks}. The network here is composed of three computer nodes (Host1-3) which are connected using an Ethernet switch. Three publishers from two different machines (Pub.1 and Pub.2 from Host1 and Pub.3 from Host2) push messages to three subscribers (Sub.1-3) in another machine (Host3). The arrows represent individual data--flow channels from the publishers to the subscribers. We want to prioritize the channel from Pub.1 to Sub.1 (bold line) so that bandwidth and resources used by the other channels do not interfere with the messages traveling from Pub.1 to Sub.1. In other words, messages from Pub.1 need to be guaranteed (in a best--effort sense) to reach Sub.1 with lowest latency. 

\subsection{Overview of YARP}
\begin{figure}[t]
   \begin{center}
     \includegraphics[width=2.8in]{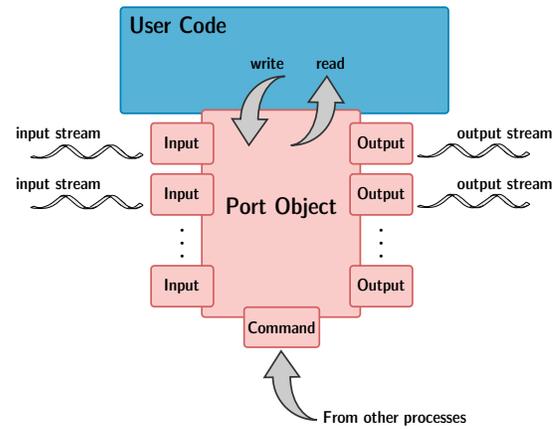}
     \caption{The internal structure of a YARP port object.}
     \label{fig:port_object}
   \end{center}
\end{figure}

YARP is a multi--platform distributed robotic middleware which consists of a set of libraries, communication protocols, and tools to keep software modules and hardware devices cleanly decoupled. Communication uses special objects called ``port''. Using ports publishers can send data to any number of receivers (subscribers), either within the same process, crossing the boundaries of processes and, using network protocols, even across machines. YARP manages connections in a way that decouples publishers and subscribers. A port is an active object that can manage multiple connections either as input or output (see Fig.~\ref{fig:port_object}); ports can be located on the network by symbolic names (e.g., \texttt{/my-port}) which are managed by a name server. The name server maps symbolic names (strings) into the triplet composed of the IP address, port number, and interface name. This information is all that is required to establish a connection between two endpoints. After such a connection is established, communication is performed by the two endpoints independently of the central server. It is worth noting that each connection has a state that can be manipulated by external (administrative) commands, which manage the connection or obtain state information from it.  

Ports can be connected using standard protocols (e.g. TCP, UDP, MCAST) either programmatically or at runtime using administrative commands. A single port may transmit the same message across several connections using different protocols; likewise it may receive messages from several sources using different protocols. Protocols can also be custom. A plug-in system allows adding ``carrier'' objects that implement new protocols. In fact the the YARP library treat connections in broad terms (is it reliable? is there a way to include meta-data? are replies possible? etc.) and delegate the actual communication to the carrier objects. Carriers have had a variety of use-cases and are a distinguishing feature of YARP. For example they may allow sending messaging over a new type of network, implement data compression or even support network-level interoperation with non-YARP programs. More recently carriers have been extended with hooks that can execute code to perform arbitrary actions locally to a component rather than remote from it~\cite{paikan13}. 

\subsection{Integration of priority QoS in YARP}
\begin{figure}[t]
   \begin{center}
     \includegraphics[width=3.4in]{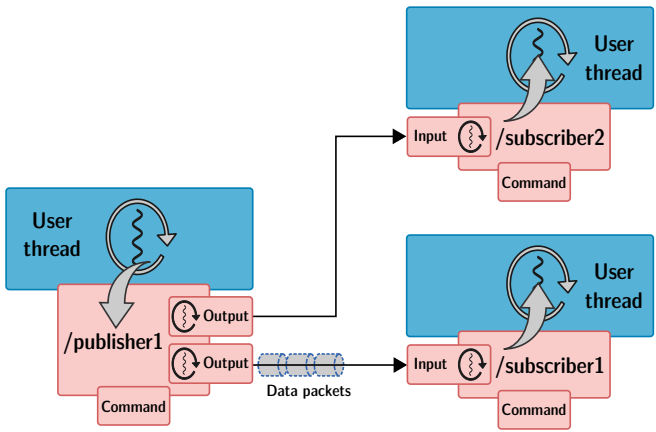}
     \caption{An example of components asynchronous communication in YARP. A publisher is pushing messages to two different subscribers using separate dedicated threads.}
     \label{fig:pubsub_example}
   \end{center}
\end{figure}

Asynchronous communication in YARP can be achieved in different ways. One way to do this in a protocol--independent manner is to configure the port object to send and receive user data in separate threads. A conceptual example is depicted in Fig.~\ref{fig:pubsub_example} in which an asynchronous publisher (Publisher 1) pushes data to two different subscribers using a separate dedicated thread for each communication channel. This decouples timing between a publisher and its subscribers and reduces the amount of time spent in the user thread for sending data. Inside the subscribers a dedicated thread read data from a communication channel. Ideally this could be avoided by relying on some OS--dependent functionalities (such as I/O signals) and notify the user thread when data is available in the subscriber port.%

Generally speaking, using dedicated threads for communication introduces extra computational time to the component execution due to thread scheduling and context--switching overhead; However, it can provide a better implementation abstraction and potentially can be exploited for the implementation of prioritized communication channels. Real--time properties and QoS attributes can be configured at different scopes, i.e., user thread, dedicated communication thread and data packets. Configuring real--time properties such as priority or scheduling policy of the user thread can be done either programmatically from the user code or automatically using component middleware functionalities and dedicated tools~\cite{mastrogiovanni2013semantic} (although this is beyond the scope of this paper).

In our approach real--time properties can be configured separately for each communication thread. In other words we modified YARP adding the possibility to change the priority of the thread that handles data transmission over a communication channel.  When a publisher writes data to a port it handles it to the corresponding thread. At this point it is the job of the operating system to schedule the threads with respect to their real--time properties so that highest--priority threads can write data to the socket before the others. 

As shown in Fig.~\ref{fig:port_object} a port object can also subscribe to multiple publishers using separate input channels. In such a case, thread prioritization can be also applied at the subscriber side to ensure that messages in a specific channel will be delivered to the user with high priority (i.e. minimum jitter).

\subsection{Data packets prioritization}

Configuring the real--time priorities of communication threads guarantees that messages are written or read from channels with specific priorities. Normally when messages are written to a socket they are handed over the operating system and there is no control on the order in which they are actually transmitted to the transport layer. Generally speaking, there is no silver bullet to data packet prioritization in computer networks. Some partial solutions exists and are highly depend on the network topology, infrastructure and communication protocol. However in Ethernet local area networks (LAN) data packets can have configurable properties that specify priority of delivery with respect to time and order.

To clarify the issue, we consider the network architecture from Fig.~\ref{fig:bottlenecks}. There are two places in our network in which packet traffic congestions can potentially happen: \emph{i)} in the OS level (i.e. inside the network driver) when outbound data from multiple applications is written to the network interface controller of the Host1, \emph{ii)} in the switch,  when packets from different ports (Host1 and 2) are forwarded to a single port (Host3). These are common bottlenecks in computer networking that become particularly critical when the infrastructure does not have enough resources for routing all traffic.

A driver queue (typically implemented as a FIFO buffer) bridges the IP stack and the network interface controller (NIC). In some operating systems (e.g., Linux) there is an intermediate layer between the IP stack and the driver queue which implement different queuing policies. This layer implements traffic management capabilities including traffic prioritization. For example in Linux distributions, the default queuing policy (i.e., pfifo\_fast QDisc)~\cite{almesberger1999linux},\cite{almesberger1999differentiated} implements a simple three band prioritization scheme based on the IP packet's TOS~\cite{almquist1992type} bits. Within each band packets follow a FIFO policy. However, prioritization happens across bands: as long as there are packets waiting in higher--priority band, the lower bands will not be processed. To hand the user--level messages to the network controller in a prioritized manner, it is enough to provide a mechanism so that the TOS bits of the data packets can be adjusted according to the desired priority of the channels.

Multilayer network switches (i.e. operating on OSI layer 3 or 4) are capable of implementing different QoS such as packet prioritization, classification and output queue congestion management. They commonly use Differentiated Services Code Point~\cite{nichols1998definition} (DSCP) which is the six most significant bits in the TOS byte to indicate the priority of an IP Packet. Differentiated services enable different classes of prioritization which can be used to provide low latency to critical network traffic while providing simple best--effort service to non--critical applications. To guarantee low--latency packet transfer from a publisher to subscriber, the TOS bits can be adjusted properly to fall into the highest--priority band of queuing policy and to form a high--priority class of differentiated service in network switches.  

\section{Implementation and results}
\label{implementation}
As described earlier port administrative commands provide a rich set of functionalities to monitor and change the state of a port and its connections. To implement the channel prioritization in YARP, these functionalities were extended to allow tuning QoS and real--time properties of port objects with the granularity of individual connections. 

In the current implementation, the port administrator provides two set of commands that affect the priority of a communication channel: setting the scheduling policy/priority of a communication thread and configuring the TOS/DSCP bits for the data packets it delivers. For example, we can simply configure real--time properties of the output entity of \texttt{/publisher1} from Fig.~\ref{fig:pubsub_example} using the YARP tools as follows: 

\lstset{captionpos=n, caption={},label=lst:thread-policy}
\begin{lstlisting}[language=Yarp, frame=single] 
 $ yarp admin rpc /publisher1
 >> prop set /subscriber1 (sched 
                          ((policy SCHED_FIFO) 
                           (priority 30))) 
\end{lstlisting}

The first line "\texttt{yarp admin rpc}" simply opens an administrative session with the port object of \texttt{/publisher1}. The second line is the real command to the administrative port. It adjusts the scheduling policy and priority of the thread in \texttt{/publisher1} which handles the connection to \texttt{/subscriber1} respectively to SCHED\_FIFO and 30 on Linux machines~\footnote{The thread scheduling properties are highly OS dependent and a proper combination of priority and policy should be used. In some platform adjusting thread priorities may not be entirely possible or is subject to specific permission.}.

For packet priorities we have chosen four predefined classes of DSCP. These classes are selected so that packets can be treated similarity by the OS queuing policy (if available) and in the network switch. For example a packet with priority class \texttt{Low} will be in the lowest priority band (Band 2) of the Linux queuing policy and will have the lowest priority in the network switch. Table~\ref{tb:class} provide a list of these classes.%

\begin{table}[ht]
\caption{Predefined classes of packet priority}
\centering
\begin{tabular}{l c c c} 
\hline\hline 
Class & DSCP & QDisc\\ [0.5ex]
\hline
\textbf{Low}      & AF11 & Band 2\\
\textbf{Normal}   & Default & Band 1\\
\textbf{High}     & AF42 & Band 0\\
\textbf{Critical} & VA & Band 0\\
\hline
\end{tabular}
\label{tb:class} 
\end{table}

Similarly data packet priority can be configured via administrative commands by setting one of the predefined priority class (or by directly configuring the DSCP/TOS bits):

\lstset{captionpos=n, caption={},label=lst:thread-policy}
\begin{lstlisting}[language=Yarp, frame=single] 
 $ yarp admin rpc /publisher1
 >> prop set /subscriber1 (qos ((priority HIGH))) 
\end{lstlisting}

This simply sets the outbound packets priority from \texttt{/publisher1} to \texttt{/subscriber1} to HIGH. 

These two set of parameters can be set for every channel in the same way and jointly define the actual priority of a communication channel in our publish/subscribe architecture.
        
\subsection{Evaluation}
\begin{figure}[t]
\centering
     \begin{subfigure}{\columnwidth}
     		\centering
           \includegraphics[width=3.2in]{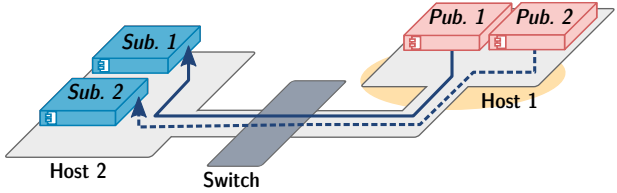}           
           \caption{Evaluating channel prioritization at the network card.}
           \label{fig:test_1}
     \end{subfigure}
     
	\vspace{0.3cm}
     \begin{subfigure}{\columnwidth}
     		\centering
           \includegraphics[width=3.3in]{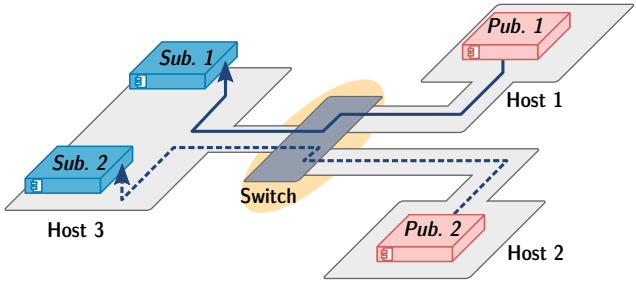}
           \caption{Evaluating channel prioritization at the network switch.} 
           \label{fig:test_2}         
	 \end{subfigure}
	 \caption{Evaluating channel prioritization QoS in YARP. For both cases, the bold line represents the prioritized channel and the dashed--line is channel which is used to produce arbitrary network load. } 
   \label{fig:test_performance}
\end{figure}
   
To evaluate our implementation of to channel prioritization in YARP, we have devised two different test cases (see Fig.~\ref{fig:test_performance}). The first case (Fig.~\ref{fig:test_1}) deals with evaluating channel prioritization when traffic congestion happens at the network card (OS level) while the second one (Fig.~\ref{fig:test_2}) investigate performance improvement due to channel prioritization at the network switch. The nodes (Host 1 to 3) are Linux machines with PREEMPT-RT kernel which are connected using Gigabit Ethernet and a QoS--enabled switch (CISCO Catalyst 2960). In both test cases there are two separate channels between the publishers and the subscribers. We measure the round--trip time of messages in the first channel (from Pub.1 to Sub.1). This is done via acknowledgment packets from Sub.1 to Pub.1 for each messages received by Pub.1. The second channel (from Pub.2 to Sub.2) produce arbitrary but controllable network load.  

\begin{figure}[t]
\centering
     \begin{subfigure}{\columnwidth}
     		\centering
		     \includegraphics[width=3.2in]{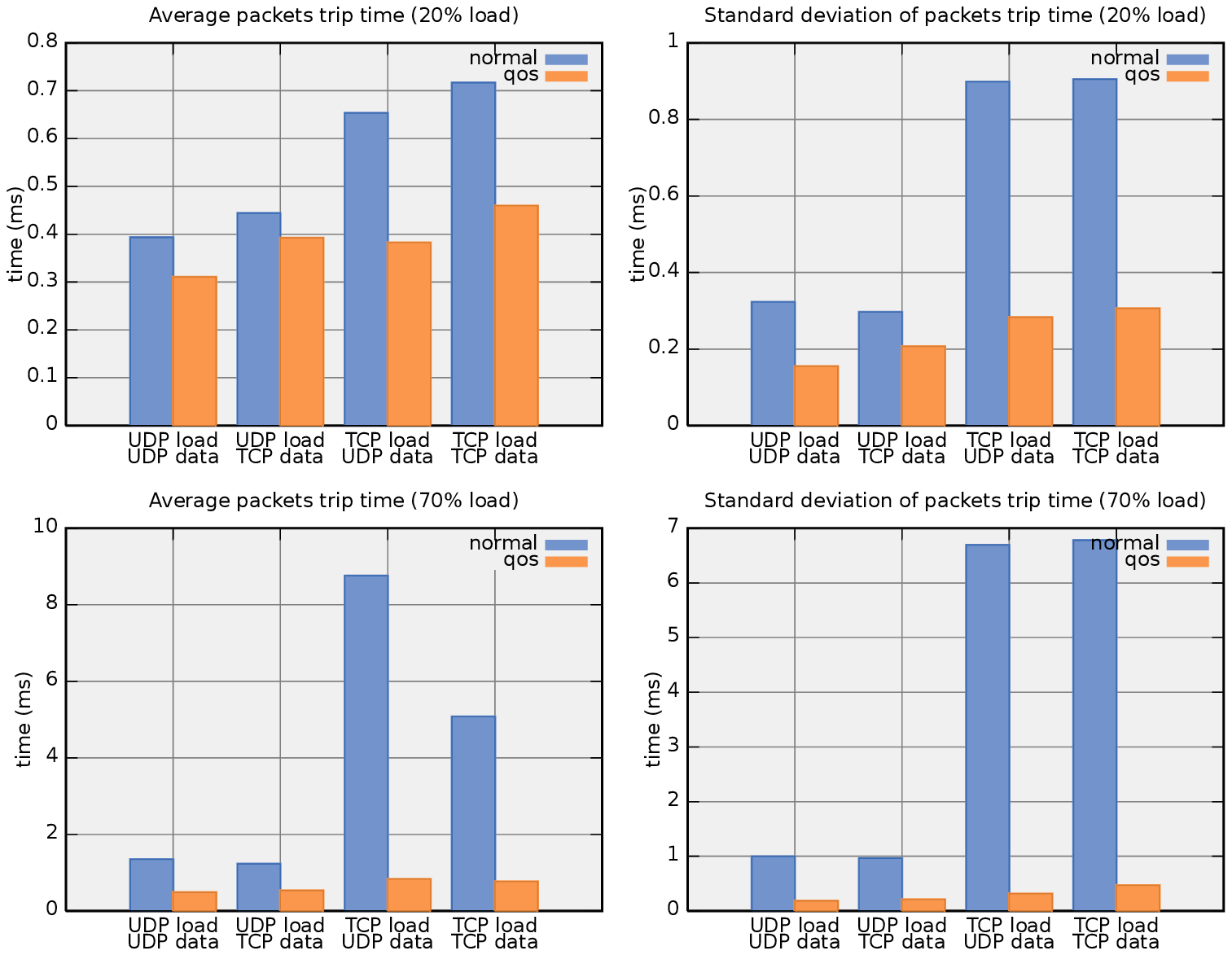}
		     \caption{Evaluating channel prioritization at the network card.}
		     \label{fig:summary_peer}  
     \end{subfigure}
     
	\vspace{0.5cm}
     \begin{subfigure}{\columnwidth}
     		\centering
	     \includegraphics[width=3.2in]{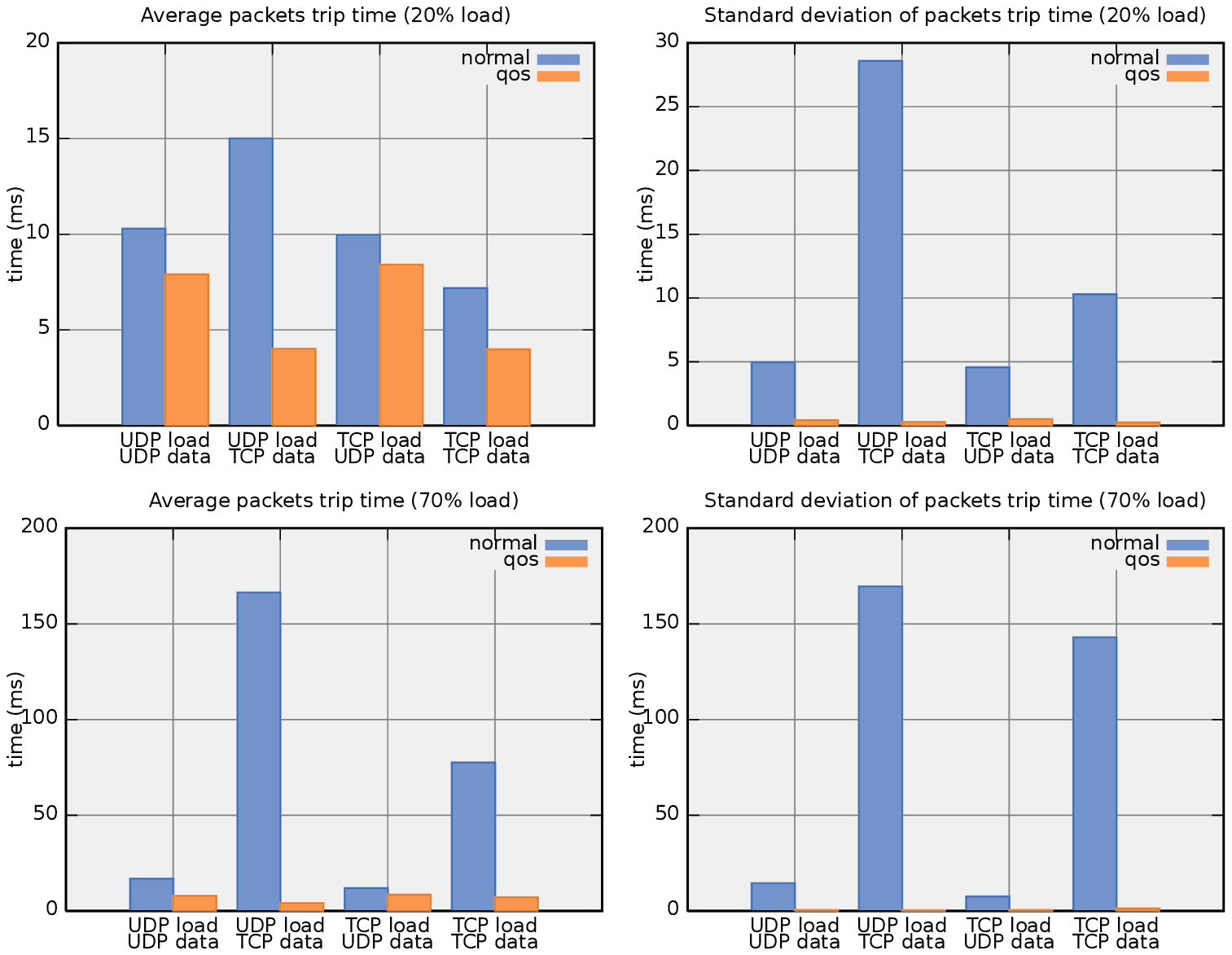}
	     \caption{Evaluating channel prioritization at the network switch.}
	     \label{fig:summary_switch}    
	 \end{subfigure}
	 \caption{Evaluation results. The bars labeled as ``qos'' represent the test results in presence of channel prioritization. The bars labeled as ``normal'' represent the test results in absence of any prioritization.} 
   \label{fig:summary}
\end{figure}

For each test case, two different set of experiments have been performed to measure packet trip time with and without channel prioritization. To achieve the highest priority for the channel from Pub.1 to Sub.1, the scheduling policy and priority of communication threads are respectively set to SCHED\_FIFO and 30. The thread priority is chosen so that it is higher than the other processes during the experiment but lower than network interrupt priorities. The packet priority for the corresponding channel is also set to HIGH (AF42/Band 0). The tests are repeated with two different network loads (corresponding respectively to 20\% and 70\% of the maximum bandwidth) generated by Pub.2 and Sub.2. Moreover, to see the effect of the underlying communication protocol on channel prioritization we have repeated the tests with different combination of TCP and UDP both for the load and the measurement channels.

Fig.~\ref{fig:summary} reports the measured average and standard deviation of the packets trip time with (bars labeled as ``qos'') and without channel prioritization (bars labeled as ``normal'') for different protocols and network loads. Fig.~\ref{fig:summary_peer} demonstrates the comparison when the two publishers (Pub.1 and Pub.2) are on the same machine and produce outbound traffic congestion at the network card only. In general, as it can be seen in Fig.~\ref{fig:summary_peer} channel prioritization produces slightly lower (in average) and more deterministic (smaller standard deviation) packet trip time. This effect is more remarkable when the network is loaded at the 70\% of the maximum bandwidth. Notice that in this case, Pub.1 and Pub.2 are located on the same machine (see Fig.~\ref{fig:test_1}) and the network load is generated by Pub.2. In this case traffic congestion happens at the level of the network card driver with consequent higher latency in packet delivery time. By prioritizing the measurement channel (bold arrow in Fig~\ref{fig:test_1}), the communication thread in Pub.1 receives higher priority by the operating system. Moreover, since packets from Pub.1 are prioritized (AF42/Band 0), they get highest priority also in the network queue and are pushed to the network physical layer before the packets from Pub.2.

Fig.~\ref{fig:summary_switch} demonstrates the results of the second test case when publishers (Pub.1 and Pub.2) are on separate machines and traffic congestion occurs at the network switch. The only difference in this case is that Pub.1 pushes larger packets to Sub.1. The reason for this is that larger data packets create higher traffic congestion in the switch. This explains why the packet trip times measured in this experiment are slightly higher than in the previous case. However, as it can be seen in Fig.~\ref{fig:summary_switch} channel prioritization greatly improved the performance (resulting in lower latency) especially when the network is highly loaded (70\%). It can also be observed a big difference in packet trip times when different communication protocols are used. The reason is that the TCP communication protocol uses the bandwidth in a smart way to achieve lower latency. Moreover, QoS--enabled network switches also have different routing policies for different packet sizes. However, as it can be seen from the results, messages transmitted through prioritized channels are comparatively less affected by differences due to the communication protocol.   

\section{Conclusions}
\label{conclusions}
This article described a novel approach to integrate priority quality of service in a publish/subscribe middleware. In our approach specific communication channels can be prioritized to ensure (in a best--effort sense) the minimum message delivery time from publishers to subscribers.  

The advantages of our approach is that it simply leverages the operating system functionalities such as real--time thread scheduling and IP packet type of service bits to deal with the typical bottlenecks that cause network congestions in local network. In addition our approach does not require centralized broker for message prioritization and for this reason it can be applied to peer-to-peer publish-subscribe architectures. Finally (although not investigated in this paper), it allows for remote and dynamic configuration of the parameters that control the communication priorities.

Our approach has been implemented in the YARP open source middleware and evaluated in two different scenarios demonstrating significant improvement in jitter and latency of message delivery, especially in presence of heavily loaded network. These results make our approach particularly useful in distributed time-critical applications. Future work will investigate mechanisms for monitoring and reconfiguring channel priorities using administrative agents and for automatically selecting optimal prioritization policies. 

\acknowledgements{
The research leading to these results has received funding from the European Union Seventh Framework
Programme [FP7-ICT-2013-10] under grant agreements n.611832, WALKMAN.
}

\bibliographystyle{abbrv}
\bibliography{citations}
\end{document}